\documentclass[aps,reprint,onecolumn, superscriptaddress, floatfix,showpacs,pre]{revtex4-2}
\usepackage{graphicx}
\usepackage{amsmath}
\usepackage{amssymb}
\usepackage{amsfonts}
\usepackage{appendix}
\usepackage{mathtools}
\usepackage{leftindex}
\usepackage{listings}

\usepackage{lipsum}

\usepackage[unicode=true,breaklinks=true,colorlinks=true]{hyperref}
\hypersetup{citecolor=blue,urlcolor=blue}

\usepackage{newtxtext}
\usepackage{newtxmath}
\usepackage[all]{hypcap}
\usepackage{natbib}
\usepackage{multirow}
\usepackage{booktabs}
\usepackage{bbold}
\usepackage{bm}
\usepackage{bbm}

\usepackage{indentfirst}
\usepackage{derivative}
\usepackage[dvipsnames]{xcolor}
\usepackage{orcidlink}
\usepackage{anyfontsize}

\newcommand{\expval}[1]{\left\langle{#1}\right\rangle}

\newcommand{\aop}{\hat{a}}
\newcommand{\adag}{\aop^\dagger}
\newcommand{\adagn}[1]{\aop^{\dagger{#1}}}

\newcommand{\comm}[2]{\left[{#1},{#2}\right]}

\newcommand{\expx}{\expval{\hat{q}}}
\newcommand{\expp}{\expval{\hat{p}}}

\newcommand{\DeleteNext}[1]{} 

\begin{document}\sloppy

\title{\texorpdfstring{Normal-ordered equivalent of the Weyl ordering of $\hat{q}^j \hat{p}^k$}{Normal-ordered equivalent of the Weyl ordering of q\^j p\^k}}

\author{Hendry M. Lim\orcidlink{0000-0002-2957-2438}}
\thanks{Current affiliation: Department of Materials Science and Engineering, National University of Singapore, Singapore 117575, Singapore}
\email[Email:~]{hendry.minfui.lim@u.nus.edu}
\affiliation{Research Center for Quantum Physics, National Research and Innovation Agency (BRIN), South Tangerang 15314, Indonesia}
\affiliation{Department of Physics, Faculty of Mathematics and Natural Sciences,
Universitas Indonesia, Depok 16424, Indonesia}

\begin{abstract}
    The problem of quantizing a bivariate dynamical system can be reduced to evaluating the ordering of $\hat{q}^j \hat{p}^k$. Here, we consider the Weyl ordering of $\hat{q}^j \hat{p}^k$ that is then expressed in term of the annihilation $\aop$ and creation $\adag$ operator. The explicit formula for the normal-ordered equivalent (all $\adag$'s preceeding all $\aop$'s) of the resulting expression is then given, and some relations are discussed.  
\end{abstract}

\maketitle

\section{Introduction}

A bivariate dynamical system is typically described in a 2-dimensional phase space by the canonical position $q$ and canonical momentum $p$, endowed with the Poisson bracket $\left\{q,p\right\}=1$. In an attempt to quantize (i.e., build a quantum mechanical description of) the system, we replace $q$ by its operator equivalent $\hat{q}$, and $p$ by $\hat{p}$, defining a Hilbert space endowed by the commutator $\comm{\hat{q}}{\hat{p}}=i\hbar$. Doing this, however, comes with a problem: quantizing terms like $qp$ introduces ordering ambiguity; we may have $\hat{q}\hat{p}$, or $\hat{p}\hat{q}$, or $\pi\hat{q}\hat{p} + (1-\pi)\hat{p}\hat{q}$, etc. The obstruction to quantization culminates in the no-go theorem of Groenewold and Van-Hove, which states that there is no quantization map that can globally preserve the Poisson bracket structure~\cite{curtright2013concise}. Nevertheless, quantization of a classical system does result in a quantum system that may reasonably mimic the classical system~\cite{BenArosh2021, Chia2020, limTransientDynamicsQuantum2025}, making the procedure worth doing in theoretical approaches. 

In this work, we consider the combinatorics from choosing the Weyl ordering during the quantization procedure~\cite{Weyl1927}. This ordering is particularly important since it corresponds to the paradigmatic Wigner representation of quantum systems in phase space~\cite{Wigner1932}, forming the so-called Wigner-Weyl correspondence. We show that the analysis of Weyl-ordered quantization for a bivariate dynamical system can be reduced to evaluating the Weyl ordering of the operator $\hat{q}^j\hat{p}^k$. Our contribution is to express the Weyl ordering in terms of the widely-used annihilation $\aop$ and creation $\adag$ operator, then rewrite the result into its normal-ordered equivalent, where all $\adag$'s precede all $\aop$'s in each term. 


\section{Discussion}

\subsection{\texorpdfstring{Quantizing a bivariate dynamical system boils down to ordering $\hat{q}^j\hat{p}^k$}{Quantizing a bivariate dynamical system boils down to ordering q\^j p\^k}}

A general bivariate classical system is defined by the equations
\begin{subequations}
\begin{align}
    q' &= A(q,p);
    \label{eq:1a}
    \\
    p' &= B(q,p),
    \label{eq:1b}
\end{align}
\end{subequations}
where the prime denotes differentiation with respect to time. We consider the case where $A$ and $B$ are polynomials in $q$ and $p$,
\begin{equation}
    A(q,p,t)=\sum_{j=0}^\infty\sum_{k=0}^\infty \mathcal{A}_{jk} q^j p^k,
\end{equation}
and likewise for $B$. Formally, quantizing this system means finding a quantum system that is \emph{expected} to generate the dynamics. The quantum expected dynamics corresponding to the classical equation of motion is obtained by replacing $q$ and $p$ with the corresponding quantum expectation values $\expval{\hat{q}}$ and $\expval{\hat{p}}$,
\begin{subequations}
\begin{align}
    \expx' &= \expval{\hat{A}}_\mathcal{W};
    \label{eq:A4a}
    \\
    \expp' &= \expval{\hat{B}}_\mathcal{W}, \label{eq:A4b}
\end{align}
\end{subequations}
where we have kept the hats inside the bra-kets to emphasize the noncommuting properties of the operators involved. In this work, we specifically consider the Weyl ordering, as indicated by the subscript $\mathcal{W}$. In Weyl-ordered quantization, an expression is mapped to the average of all possible operator orderings; for example, $qp$ is mapped to ${\left(\hat{q}\hat{p} + \hat{p}\hat{q}\right)}/{2}$~\cite{hall2013quantum}. We have \begin{equation}\label{eq:A5}
    \expval{\hat{A}}_\mathcal{W} = \sum_{j=0}^\infty\sum_{k=0}^\infty\frac{\mathcal{A}_{jk}}{N_{jk}}\expval{\sum_\omega \left(\hat{q}^j\hat{p}^k\right)_\omega},
\end{equation}
and likewise for $\expval{\hat{B}}_\mathcal{W}$. Here $N_{jk}=(j+k)!/j!k!$ is the number of possible orderings $\omega$ of $\hat{q}^j\hat{p}^k$. We can treat each term in the series on its own, whence $N_{jk}^{-1} \sum_\omega \left(\hat{q}^j\hat{p}^k\right)_\omega$ is just the Weyl ordering of $\hat{q}^j \hat{p}^k$.  


\subsection{The ladder operators}

Typically, it is practical to introduce the annihilation $\aop$ and creation $\adag$ operators by
\begin{equation}
    \aop = \frac{\hat{q}+i\hat{p}}{\sqrt{2\hbar}},
\end{equation}
satisfying $\comm{\aop}{\adag}=1$. These operators are useful as many quantum mechanical models can be described as modifications to the quantum simple harmonic oscillator~\cite{BenArosh2021, Chia2020, limTransientDynamicsQuantum2025}, for which $\aop$ annihilates one quantum of energy, while $\adag$ creates one quantum of energy. Moreover, the quantization procedure is more convenient using these operators~\cite{Cahill1969, Cahill1969.II, lim2025algebraicmachineryquantization}. In terms of these operators, we can define the normal ordering and antinormal ordering. With the former, all $\adag$'s precede all $\aop$; with the latter, we have the opposite~\cite{hall2013quantum}. The normal ordering is particularly useful since any expression in this ordering is physically measurable (for example, see Section 3.7 of Ref.~\cite{gerry2005introductory}). Based on this understanding, we are motivated to rewrite the quantum expected dynamics in terms of $\aop$ and $\adag$, then evaluate the normal-ordered equivalent of the result. 

\subsection{Main result}

Instead of summing over all possible orderings $\omega$, let us sum over all possible ``forced orderings'' $\sigma$ of $\hat{q}^j\hat{p}^k$. By forced ordering, we mean swapping two identical operators and considering it as a separate ordering. For example, the forced ordering of $\hat{q}^2\hat{p}$ gives us $\hat{q}\hat{q}\hat{p},\hat{q}\hat{q}\hat{p},\hat{q}\hat{p}\hat{q},\hat{q}\hat{p}\hat{q}, \hat{p}\hat{q}\hat{q},\hat{p}\hat{q}\hat{q}$. As shown below, the moments appearing in the final result depend on $(j+k)$, not their individual values. Forced ordering is thus more convenient, as the number of orderings does not depend on the individual values of $j$ and $k$. For a given $(j,k)$, we consider the average over the forced orderings $\sigma$ of the operators%
~\footnote{If we assume not to know what values $s_{\sigma(r)}^{(jk)}$ may take, then we can think of the sum over $\sigma$ as the sum over all permutations of $\left\{s_{\sigma(r)}^{(jk)}\right\}$---hence our choice of notation.}%
,
\begin{equation}
\begin{split}
    \hat{S}_{jk}&=\frac{1}{(j+k)!}\sum_\sigma \left(\hat{q}^j\hat{p}^k\right)_\sigma
    \\
    &=\frac{i^k}{2^{(j+k)/2}(j+k)!}\sum_\sigma \prod_{r=1}^{j+k}\left(\adag+s_{\sigma(r)}^{(jk)}\aop\right),
\end{split}
\end{equation}
where $s_{\sigma(r)}^{(jk)}$ is the sign of the $r$th factor for the forced ordering $\sigma$ of $\hat{q}^j\hat{p}^k$. If the corresponding operator is $\hat{q}$, then $s_{\sigma(r)}^{(jk)}=1$; otherwise, it equals $-1$. Evaluating the product, we obtain the general form
\begin{equation}\label{eq:S_jk}
\begin{split}
    \hat{S}_{jk}&=\frac{i^k}{2^{(j+k)/2}(j+k)!}
      \sum_\sigma \sum_{u=0}^{(j+k)/2}\sum_{v=0}^{j+k-2u}\eta_{jkuv\sigma}\adagn{(j+k-2u-v)}\aop^v.
\end{split}
\end{equation}
Here $\eta_{jkuv\sigma}$ is a polynomial in $s^{(jk)}_{\sigma(r_1)}, s^{(jk)}_{\sigma(r_2)}, \dots, s^{(jk)}_{\sigma
(r_{u+v})}$, where $r_{1},r_2,\dots r_{u+v}$ denote some $(u+v)$ of the $(j+k)$ signs. For example, with $j+k=3$, we have $\eta_{jk11\sigma} = 2s_{\sigma(1)}s_{\sigma(2)}+s_{\sigma(1)}s_{\sigma(3)}$. To arrive at Eq.~\eqref{eq:S_jk}, we have utilized the commutation relation $\comm{\aop}{\adag}=1$. We find looking for the general form of $\eta_{jkuv\sigma}$ difficult, and instead consider the sum over $\sigma$ for a given $(u,v)$. We evaluate the sum with mathematical induction and find that the sum over $\sigma$ may be written as a product of three quantities:
\begin{equation}\label{eq:eta_jkpq_sigma}
    \sum_{\sigma}\eta_{jkuv\sigma} = \lambda_{jkuv}\xi_{jkuv}\zeta_{jkuv}.
\end{equation}
When we write out the sum, the term $s^{(jk)}_{\sigma(r_1)}s^{(jk)}_{\sigma(r_2)}\dots s^{(jk)}_{\sigma(r_{u+v})}$ for some $\sigma$ appears exactly $\lambda_{jkuv}$ times in the place of each term of $\eta_{jkuv\sigma}$. In general,
\begin{equation}
\begin{split}
    \lambda_{jkuv} &= (j+k-u-v)!(u+v)!.
\end{split}
\end{equation}
Summing like terms and factoring out $\lambda_{jkuv}$, we have the sum $\xi_{jkuv}$ of all coefficients in $\eta_{jkuv\sigma}$, generally given by the concatenated Bessel-scaled Pascal triangles~\cite{oeis}
\begin{equation}
    \xi_{jkuv} = \frac{(j+k)!}{2^u u! v! (j+k-2u-v)!}.
\end{equation}
The quantity $\lambda_{jkuv}\xi_{jkuv}$ is the total coefficient of each distinct term. Summing over the distinct terms and factoring out the common $\lambda_{jkuv}\xi_{jkuv}$, we have
\begin{subequations}
\begin{align}
    \zeta_{jkuv}&=\sum_{r_1=1}^{j+k}\sum_{r_2=r_1+1}^{j+k}\dots\sum_{r_{u+v}=r_{u+v-1}+1}^{j+k} s_{r_1}^{(jk)} s_{r_2}^{(jk)} \dots s_{r_{u+v}}^{(jk)};
    \\
    \zeta_{jk00}&=1,
\end{align}
\end{subequations}
where $s_{r_m}^{(jk)}$ can be of \emph{any} $\sigma$ since we may start with any $\sigma$ and sum through all the permutations. We find it convenient to choose $\sigma$ such that $S_{r_m}^{(jk)}=1$ for $r_m=1,2,\dots,j$ and $-1$ otherwise. We obtain
\begin{equation}
\begin{split}
    \zeta_{jkuv} &=\sum_{m=0}^{u+v}(-1)^m \binom{j}{u+v-m}\binom{k}{m},
\end{split}
\end{equation}
where $\binom{a}{b}$ is the binomial coefficient we take to be zero when $a<b$ or $b<0$. This summation is known to be the alternating-sign Vandermonde convolution, whose alternative expression is~\cite{vandermonde}
\begin{equation}\label{eq:zeta_jkpq}
    \zeta_{jkuv} = \left[x^{u+v}\right]\left(1+x\right)^j\left(1-x\right)^k,
\end{equation}
i.e., the coefficient of $x^{u+v}$ in the polynomial $(1+x)^j(1-x)^k$. Substituting Eqs.~\eqref{eq:eta_jkpq_sigma}--\eqref{eq:zeta_jkpq} into Eq.~\eqref{eq:S_jk}, we finally obtain
\begin{subequations}
\begin{align}\label{eq:expected_phase_point_dynamics}
    \hat{S}_{jk} &= \sum_{u=0}^{(j+k)/2}\sum_{v=0}^{j+k-2u} h_{jkuv}\adagn{(j+k-2u-v)}\aop^v;
    \\
    \label{eq:h_jkpq}
    h_{jkuv}&=\frac{i^k}{2^{(j+k)/2}}\frac{u!}{2^u}\binom{[j+k]-[u+v]}{u}\binom{u+v}{u}\left\{\left[x^{u+v}\right]\left(1+x\right)^j\left(1-x\right)^k\right\}.
\end{align}
\end{subequations}
We restate here that $\hat{S}_{jk}$ is the Weyl ordering of $\hat{q}^j\hat{p}^k$, and the above expression gives its normal-ordered equivalent. 


\subsection{Symmetry}

The coefficients $h_{jkuv}$ satisfy
\begin{subequations}
\begin{align}
    h_{jkuv} &= (-1)^k h_{jku(j+k-2u-v)};
    \label{eq:simultaneity_of_term_pairs}
    \\[5pt]
    h_{jku([j+k-2u]/2)} &= 0 \quad \text{if $j$ and $k$ are odd.}
    \label{eq:zero_terms_for_odd_jk}
\end{align}
\end{subequations}
One can show that the product of the factors excluding $\left[x^{u+v}\right] (1+x)^j(1-x)^k$ is identical for both $h$ coefficients in both equations, whence the properties are consequences of how the polynomial $(1+x)^j(1-x)^k$ behaves under $x\mapsto -x$. One may argue that these properties are rather trivial, since they translate to the requirement that ${\adagn{(j+k-2u-v)}\aop^v}$ and ${\adagn{v}\aop^{j+k-2u-v}}$ form a conjugate transpose pair; accounting for the factors $i^k$ and $(-1)^k$, we can see that the requirement arises because we are quantizing a real-valued expression. 


\section{Concluding Remarks}

We have evaluated the normal-ordered equivalent of the Weyl ordering of $\hat{q}^j\hat{p}^k$ using a brute force approach (see also the Supplementary Materials). In the following, we note several other methods to tackle this problem, which should reproduce our end result or give an alternative expression. 

Cahill and Glauber~\cite{Cahill1969} formulated an explicit formula to write an unevaluated Weyl ordering of $\aop^m\adagn{n}$ in terms of normal ordered expressions. Let $\left\{\aop^m\adagn{n}\right\}_\mathcal{W}$ be the Weyl ordering of $\aop^m\adagn{n}$ (e.g., $\left\{\aop\adag\right\}_\mathcal{W} = {\left(\aop\adag+\adag\aop\right)}/{2}$). Then,
\begin{equation}
    \left\{\aop^m\adagn{n}\right\}_\mathcal{W} = \sum_{l=0}^{\min(m,n)} \frac{l!}{2^l} \binom{m}{l} \binom{n}{l}  \adagn{(n-l)}\aop^{(m-l)}.
\end{equation}
One can thus first write the Weyl ordering of $\hat{q}^j\hat{p}^k$ in terms of $\left\{\aop^m\adagn{n}\right\}_\mathcal{W}$ before using the above equation. 

Furthermore, Blasiak~\cite{blasiak2005combinatoricsbosonnormalordering} formulated an explicit formula for the normal-ordered equivalent of any string built out of $\aop$ and $\adag$, called a ``boson string''. Let
\begin{equation}
    \hat{X} = \adagn{r_M}\aop^{s_M}\adagn{r_{M-1}}\aop^{s_{M-1}} \dots  \adagn{r_2}\aop^{s_2}\adagn{r_1}\aop^{s_1}
\end{equation}
be a boson string. Then, its normal-ordered equivalent is given by
\begin{equation}
    \hat{X} = \begin{cases}
    \displaystyle
        \sum_{k=s_1}^{s_1+s_2+\dots+s_M} S_{\bm{r},\bm{s}}(k) \adagn{(d_M+k)}\aop^k, & d_M \geq 0;
        \\
        \displaystyle
        \sum_{k=r_M}^{r_1+r_2+\dots+r_M} S_{\overline{\bm{s}}, \overline{\bm{r}}}(k) \adagn{k}\aop^{(-d_M+k)}, & d_M < 0,
    \end{cases}
\end{equation}
where $\bm{r}=(r_1,r_2,\dots, r_M)$, $\bm{s} = (s_1, s_2, \dots, s_M)$, $\overline{\bm{r}} = (r_M, r_{M-1},\dots, r_1)$, $\overline{\bm{s}}=(s_M, s_{M-1},\dots, s_1)$,
\begin{equation}
    d_l = \sum_{m=1}^l (r_m-s_m),
\end{equation}
and 
\begin{equation}
    S_{\bm{r},\bm{s}}(k) = \frac{1}{k!} \sum_{j=0}^k \binom{k}{j} (-1)^{k-j} \prod_{m=1}^M \left(d_{m-1}+j\right)_{s_m},
\end{equation}
where $(m)_n = m! / (m-n)!$ is the falling factorial. One may rewrite the Weyl ordering of $\hat{q}^j \hat{p}^k$ in terms of $\aop$ and $\adag$, and use Blasiak's formula above to evaluate its normal-ordered equivalent. 

\bibliography{refs}

\clearpage

\section*{Supplementary Materials}

\subsection{\texorpdfstring{Obtaining the general form of $\lambda_{klpq}$ and $\xi_{jkpq}$}{Obtaining the general form of lambda\_klpq and xi\_jkpq}}\label{section_obtaining_the_general_form_of}

As an example, let us consider the expansion for $j+k=3$ for a given permutation $\sigma$ (dropping their corresponding notations in the equation):
\begin{equation}
\begin{split}
    \prod_{r=1}^3\left(\adag+s_r\aop\right) &=\adagn{3}+(s_1+s_2+s_3)\adagn{2}\aop+(s_1s_2+s_1s_3+s_2s_3)\adag\aop^2+ s_1s_2s_3\aop^3
    \\
    &\quad + (2s_1+s_2)\adag+(2s_1s_2+s_1s_3)\aop
\end{split}
\end{equation}
We can immediately see the pattern where all the moments are of orders $j+k$ (first row), then $j+k-2$ (second row), $j+k-4$ (does not apply here), $\dots$, $0$ if $j+k$ is even and $1$ if odd. The terms have been grouped by $p$ and $q$. We have $(p,q)=(0,0),(0,1),(0,2),(0,3),(1,0),(1,1)$, in that order. We are interested in the sum over all permutations $\sigma$ for the given $j,k,p,q$. As a less obvious example, let us pick the term with $\aop$. Let the expression shown, $(2s_1s_2+s_1s_3)\aop$, be the $(1,2,3)$ permutation. Then, we have five other permutations to write out. The sum is given by (explicitly writing $0 s_{\sigma(2)}s_{\sigma(3)}$ so we have all the possible $(p+q)$ out of $(j+k)$ signs)
\begin{equation}
\begin{split}
    \sum_\sigma 2s_{\sigma(1)}s_{\sigma(2)}+s_{\sigma(1)}s_{\sigma(3)}+0s_{\sigma(2)}s_{\sigma(3)} &= (2s_1s_2+s_1s_3+0s_2s_3) +(2s_1s_2+s_1s_2+0s_3s_2)
    \\
    &\quad+(2s_2s_1+s_2s_3+0s_1s_3)+(2s_2s_3+s_2s_1+0s_3s_1)
    \\
    &\quad+(2s_3s_1+s_3s_2+0s_1s_2)+(2s_3s_2+s_3s_1+0s_2s_1)
\end{split}
\end{equation}
Notice how $s_1s_2=s_2s_1$ appears twice in every ``slot''. In the first row, it appears in the first slot; in the second to the third row and first column, it appears in the second slot; in the second to the third row and second column, it appears in the third slot. It is similar for $s_1s_3$ and $s_2s_3$. As such, \emph{$s^{(jk)}_{\sigma(r_1)}\dots s^{(jk)}_{\sigma(r_{p+q})}$ appears a constant number of times in each slot of the $\eta_{jkpq\sigma}$}. Let this number be $\lambda_{jkpq}$. If we now sum like terms, we get
\begin{equation}
\begin{split}
    \sum_\sigma 2s_{\sigma(1)}s_{\sigma(2)}+s_{\sigma(1)}s_{\sigma(3)}+0s_{\sigma(2)}s_{\sigma(3)} &= [(2)(2)+(2)(1)+(2)(0)]s_1s_2
    \\
    &\quad +[(2)(2)+(2)(1)+(2)(0)]s_1s_3
    \\
    &\quad +[(2)(2)+(2)(1)+(2)(0)]s_2s_3
    \\
    &= 2(2+1+0)(s_1s_2+s_1s_3+s_2s_3)
\end{split}
\end{equation}
Each of $s^{(jk)}_{\sigma(r_1)}\dots s^{(jk)}_{\sigma(r_{p+q})}$ appears twice in the first slot with ``weight'' 2, twice in the second slot with weight 1, and twice in the third slot with weight 0. If we factor out the common $\lambda_{jkpq}=2$, we have a sum over the weights, which are just the coefficients of $\eta_{jkpq\sigma}$. This is shared by every possible $s^{(jk)}_{\sigma(r_1)}\dots s^{(jk)}_{\sigma(r_{p+q})}$, so we can factor out the common sum of weights, which we denote $\xi_{jkpq}$, leaving out a sum over all the $s^{(jk)}_{\sigma(r_1)}\dots s^{(jk)}_{\sigma(r_{p+q})}$, which we denote $\zeta_{jkpq}$. We find that
\begin{equation}
\begin{split}
    \lambda_{jkpq} = \frac{(j+k)!}{C(j+k,p+q)}
\end{split}
\end{equation}
where $(j+k)!$ is the number of possible ways we can rearrange $s_1,s_2,\dots,s_{j+k}$ and $C(j+k,p+q)$ is the possible product-of-$(p+q)$-out-of-$(j+k)$-signs (i.e. the number of possible terms in the polynomial $\eta_{jkpq\sigma}$) there is. Meanwhile, plugging the sum-of-coefficients pattern for different $(p,q)$ into OEIS, we find that it is given by the concatenated Bessel-scaled Pascal triangle~\cite{oeis},
\begin{equation}
\begin{split}
    \xi_{jkpq} = \frac{(j+k)!}{2^p p! q! (j+k-2p-q)!}
\end{split}
\end{equation}

\clearpage
\subsection{\texorpdfstring{Evaluating $\zeta_{jkpq}$}{Evaluating zeta\_jkpq}}\label{section_evaluating_zeta_jkpq}

We to calculate the sum over all $s^{(jk)}_{\sigma(r_1)}\dots s^{(jk)}_{\sigma(r_{p+q})}$. This is a sum over products of $(p+q)$ signs. The order does not matter, as we can start with any order and write out all the permutations anyway. Generalizing from the sum $s_1s_2+s_1s_3+s_2s_3$ in our example, we have the general form
\begin{equation}
    \zeta_{jkpq} = \sum_{r_1}^{j+k}\sum_{r_2=r_1+1}^{j+k}\dots\sum_{r_{p+q}=r_{p+q-1}+1}^{j+k}s_{r_1}s_{r_2}\dots s_{r_{p+q}}
\end{equation}
Since the order of signs does not matter, let us choose
\begin{equation}
    s_p =
    \begin{cases}
    1, & p = 1,2,\dots,j\\
    -1, & p = j+1,j+2,\dots,j+k
    \end{cases}
\end{equation}
Our strategy is to break up the summation into intervals where each $s_p$ takes the same value. Doing this, however, requires that the terms exist in the first place. Here's a helpful function:
\begin{equation}
    g(a,b)=\begin{cases}
        1, & \mathrm{if}\ a\geq b
        \\
        0, & \mathrm{otherwise}
    \end{cases}
\end{equation}
As seen below, the terms resulting from the partition make it all the way to the end, even though it must be zero for the given $(j,k)$ and $p+q$. For example, with $(j,k)=(2,1)$, the third summation in ``Case $p+q=2$'' should not contribute to the sum, yet it is there at the end. Using the function $g$, we keep track of which term may contribute to the sum by ensuring that there is \textit{at least one} term in the sum partition---only then is the contribution valid. Let the ``completely partitioned sum'' be the series of sums where all $s_p$ has taken their values. This is where we can most easily ``modify'' the sum by putting in $g(a,b)$.

\vspace*{2pt}
\newcommand{\sump}[2]{\sum_{P={#1}}^{#2}}
\newcommand{\sumq}[2]{\sum_{Q={#1}}^{#2}}
\newcommand{\sumr}[2]{\sum_{R={#1}}^{#2}}
\newcommand{\sumt}[2]{\sum_{T={#1}}^{#2}}
\newcommand{\sumu}[2]{\sum_{U={#1}}^{#2}}

\noindent\textbf{Case $p+q=0$}

Here we have unity. 

\noindent\textbf{Case $p+q=1$}
\begin{equation}
\begin{split}
    \sum_{P=1}^{j+k}s_P &= g(j,1)\sum_{P=1}^{j}(1)+g(k,1)\sum_{P=j+1}^{j+k}(-1)
    \\
    &= g(j,1)j - g(k,1)k
\end{split}
\end{equation}

\noindent\textbf{Case $p+q=2$}

\begin{equation}
\begin{split}
    \sump{1}{j+k}\sumq{P+1}{j+k}s_Ps_Q
    &= \sump{1}{j}\sumq{P+1}{j+k}(1)s_Q+\sump{j+1}{j+k}\sumq{P+1}{j+k}(-1)s_Q
    \\
    &=g(j,2)\sump{1}{j}\sumq{P+1}{j}(1)(1)+g(j,1)g(k,1)\sump{1}{j}\sumq{j+1}{j+k}(1)(-1)+g(k,2)\sump{j+1}{j+k}\sumq{P+1}{j+k}(-1)(-1)
    \\
    &= g(j,2)\frac{j(j-1)}{2}-g(j,1)g(k,1)jk+g(k,2)\frac{k(k-1)}{2}
\end{split}
\end{equation}

\newpage
\noindent\textbf{Case $p+q=3$}

\begin{equation}
\begin{split}
    \sump{1}{j+k}\sumq{P+1}{j+k}\sumr{Q+1}{j+k}s_Ps_Qs_R
    &= \sump{1}{j}\sumq{P+1}{j+k}\sumr{Q+1}{j+k}(1)s_Qs_R + \sump{j+1}{j+k}\sumq{P+1}{j+k}\sumr{Q+1}{j+k}(-1)s_Qs_R
    \\
    &= \sump{1}{j}\sumq{P+1}{j}\sumr{Q+1}{j+k}(1)(1)s_R+\sump{1}{j}\sumq{j+1}{j+k}\sumr{Q+1}{j+k}(1)(-1)s_R
    \\
    &\quad + \sump{j+1}{j+k}\sumq{P+1}{j+k}\sumr{Q+1}{j+k}(-1)(-1)s_R
    \\
    &= g(j,3)\sump{1}{j}\sumq{P+1}{j}\sumr{Q+1}{j}(1)(1)(1) + g(j,2)g(k,1)\sump{1}{j}\sumq{P+1}{j}\sumr{j+1}{j+k}(1)(1)(-1)
    \\
    &\quad +g(j,1)g(k,2)\sump{1}{j}\sumq{j+1}{j+k}\sumr{Q+1}{j+k}(1)(-1)(-1) + g(k,3)\sump{j+1}{j+k}\sumq{P+1}{j+k}\sumr{Q+1}{j+k}(-1)(-1)(-1)
    \\
    &= g(j,3)\frac{j(j-1)(j-2)}{6}-g(j,2)g(k,1)\frac{j(j-1)k}{2}\\&\quad+g(j,1)g(k,2)\frac{jk(k-1)}{2}-g(k,3)\frac{k(k-1)(k-2)}{6}
\end{split}
\end{equation}

\noindent\textbf{Case $p+q=4$}

\begin{equation}
\begin{split}
    \sump{1}{j+k}\sumq{P+1}{j+k}\sumr{Q+1}{j+k}\sumt{R+1}{j+k}s_P s_Q s_R s_T
    &= g(j,4)\sump{1}{j}\sumq{P+1}{j}\sumr{Q+1}{j}\sumt{R+1}{j}(1)(1)(1)(1) 
    \\
    &\quad + g(j,3)g(k,1)\sump{1}{j}\sumq{P+1}{j}\sumr{Q+1}{j}\sumt{j+1}{j+k}(1)(1)(1)(-1)
    \\
    &\quad + g(j,2)g(k,2)\sump{1}{j}\sumq{P+1}{j}\sumr{j+1}{j+k}\sumt{r+1}{j+k}(1)(1)(-1)(-1)
    \\
    &\quad + g(j,1)g(k,3)\sump{1}{j}\sumq{j+1}{j+k}\sumr{Q+1}{j+k}\sumt{r+1}{j+k}(1)(-1)(-1)(-1)
    \\
    &\quad +g(k,4)\sump{j+1}{j+k}\sumq{P+1}{j+k}\sumr{Q+1}{j+k}\sumt{R+1}{j+k}(-1)(-1)(-1)(-1)
    \\
    &= g(j,4)\frac{j(j-1)(j-2)(j-3)}{24}-g(j,3)g(k,1)\frac{j(j-1)(j-2)k}{6}
    \\
    &\quad +g(j,2)g(k,2)\frac{j(j-1)k(k-1)}{4}
    \\
    &\quad -g(j,1)g(k,3)\frac{jk(k-1)(k-2)}{6}+g(k,4)\frac{k(k-1)(k-2)(k-3)}{24}
\end{split}
\end{equation}

\newpage
\noindent\textbf{Case $p+q=5$}

\begin{equation}
\begin{split}
    \sump{1}{j+k}\sumq{P+1}{j+k}\sumr{Q+1}{j+k}\sumt{R+1}{j+k}\sumu{T+1}{j+k}s_P s_Q s_R s_T s_U 
    &= g(j,5)\sump{1}{j}\sumq{P+1}{j}\sumr{Q+1}{j}\sumt{R+1}{j}\sumu{T+1}{j}(1)(1)(1)(1)(1)
    \\
    &\quad + g(j,4)g(k,1)\sump{1}{j}\sumq{P+1}{j}\sumr{Q+1}{j}\sumt{R+1}{j}\sumu{j+1}{j+k}(1)(1)(1)(1)(-1)
    \\
    &\quad + g(j,3)g(k,2)\sump{1}{j}\sumq{P+1}{j}\sumr{Q+1}{j}\sumt{j+1}{j+k}\sumu{T+1}{j+k}(1)(1)(1)(-1)(-1)
    \\
    &\quad + g(j,2)g(k,3)\sump{1}{j}\sumq{P+1}{j}\sumr{j+1}{j+k}\sumt{R+1}{j+k}\sumu{T+1}{j+k}(1)(1)(-1)(-1)(-1)
    \\
    &\quad + g(j,1)g(k,4)\sump{1}{j}\sumq{j+1}{j+k}\sumr{Q+1}{j+k}\sumt{R+1}{j+k}\sumu{T+1}{j+k}(1)(-1)(-1)(-1)(-1)
    \\
    &\quad + g(k,5)\sump{j+1}{j+k}\sumq{P+1}{j+k}\sumr{Q+1}{j+k}\sumt{R+1}{j+k}\sumu{T+1}{j+k}(-1)(-1)(-1)(-1)(-1)
    \\
    &=g(j,5)g(k,0)\frac{j!k!}{5!(j-5)!0!(k-0)!}-g(j,4)g(k,1)\frac{j!k!}{4!(j-4)!1!(k-1)!}
    \\
    &\quad +g(j,3)g(k,2)\frac{j!k!}{3!(j-3)!2!(k-2)!}-g(j,2)g(k,3)\frac{j!k!}{2!(j-2)!3!(k-3)!}
    \\
    &\quad +g(j,1)g(k,4)\frac{j!k!}{1!(j-1)!4!(k-4)!}-g(j,0)g(k,5)\frac{j!k!}{0!(j-0)!5!(k-5)!}
\end{split}
\end{equation}

\noindent\textbf{General formula}

\begin{equation}
\begin{split}
    \zeta_{jkpq} &= \sum_{r_1=1}^{j+k}\sum_{r_2=r_1+1}^{j+k}\dots\sum_{r_{p+q}=r_{p+q-1}+1}^{j+k} s_{r_1} s_{r_2} \dots s_{r_{p+q}}
    \\
    &= \sum_{m=0}^{p+q}(-1)^m\frac{g(j,p+q-m)j!}{(p+q-m)!(j-p-q+m)!}\frac{g(k,m)k!}{m!(k-m)!}
    \\
    &= \sum_{m=0}^{p+q}(-1)^mg(j,p+q-m) C(j,p+q-m) g(k,m) C(k,m)
\end{split}
\end{equation}
The $m$th term is zero whenever the difference between the first and second argument in the binomial coefficient is less than zero, implying that we can simplify this formula as
\begin{equation}
    \zeta_{jkpq} = \sum_{m=0}^{p+q}(-1)^m \gamma(j,p+q-m)\gamma(k,m)
\end{equation}
where
\begin{equation}
    \gamma(a,b)=g(a,b)C(a,b)=\begin{cases}
        C(a,b), & \text{if $a\geq b$}
        \\
        0, & \text{otherwise}
    \end{cases}
\end{equation}
We can also write the result in terms of nonzero contributions only, though the summation limit is not as pretty:
\begin{equation}
    \zeta_{jkpq} = \sum_{m=\max(0,p+q-j)}^{\min(k,p+q)}(-1)^m C(j,p+q-m)C(k,m)
\end{equation}
where $p+q-j$ gives the lowest $m$ for which $\gamma(j,p+q-m)=C(j,p+q-m)$ (the second argument decreases as $m$ increases) and $k$ gives the highest $m$ for which $\gamma(k,m)=C(k,m)$ (the second argument increases with $m$).

\end{document}